\documentclass[range]{ar2e}

\usepackage{latexsym}
\usepackage{amssymb,amsmath}
\usepackage{hyperref}
\input psfig.sty

\jname{Annual Review of Nuclear and Particle Science}
\jyear{(2012)}  
\jvol{62}

\newcommand{\refl}[1]{(\ref{#1})}

\newcommand{\beq}{\begin{equation}}                                             
\newcommand{\eeq}{\end{equation}}     
\newcommand{\eeql}[1]{\label{#1}\eeq}
\def\bqa{\begin{eqnarray}}                                                      
\def\eqa{\end{eqnarray}}

\newcommand{\rt}{\ensuremath{\sqrt{2}}}                                          
\newcommand{\oh}{\ensuremath{\frac{1}{2}}}                                          

\newcommand{\ra}{\ensuremath{\rightarrow}}
\newcommand{\Ra}{\ensuremath{\Rightarrow}}

\newcommand{\ord}[1]{\ensuremath{\mathcal{O}(#1)}}

\newcommand{\x}{\ensuremath{\times}}                                                
   
\newcommand{\mcal}{\ensuremath{\mathcal{M}}}                                

\newcommand{\mred}{\ensuremath{\overline{M}_P}}

\newcommand{\veva}[1]{\ensuremath{\langle#1\rangle}}

\newcommand{\lag}{\ensuremath{\mathcal{L}}}

\newcommand{\douba}[2]{\ensuremath{                                                  
\left( \begin{array}{c} #1    \\ #2 
        \end{array}\right)}}

\newcommand{\zpr}{\ensuremath{Z'}}

\newcommand{\uprm}{\ensuremath{U(1)'}}

\newcommand{\sto}{\ensuremath{SU(2) \x U(1)}}                                       
\newcommand{\st}{\ensuremath{SU(2)}}

\begin{document}
%


\title{Neutrino Masses from the Top Down}

\markboth{Paul Langacker}{Neutrino Masses from the Top Down}

\author{Paul Langacker
\affiliation{Institute for Advanced Study,
Princeton, NJ 08540, USA \\
Department of Physics, Princeton University, Princeton, NJ 08544, USA}
}


\begin{abstract}
General classes of mechanisms for generating small neutrino masses are surveyed from a top-down (superstring) perspective. In particular, string constructions have motivated various possibilities involving higher-dimensional operators, string instantons, and wave function overlaps in large or warped extra dimensions. These may yield small Dirac masses, Majorana masses via the Weinberg operator, or Majorana masses from a seesaw mechanism, though the latter typically differ in detail from the more conventional GUT models. Possibilities for mixing between light active and sterile neutrinos are surveyed.
\end{abstract}

\maketitle

\section{INTRODUCTION}
Neutrinos have played an important role in probing or testing the standard electroweak model, QCD,   the structure of the nucleon,
the interior of the Sun, the dynamics of core collapse supernovae and other violent astrophysical events, and recently even the interior of the Earth. The neutrino masses are especially intriguing, even compared to the (also not understood) quark and lepton masses:
they are much lighter, two of the three mixing angles are large, they could be either Dirac or Majorana, and there are even hints of mixing between active and sterile neutrinos. The small neutrino masses could be inversely proportional to large new physics scales, as in superstring theories or grand unification, or they could be associated with physics at the TeV or intermediate scales. The physics responsible for neutrino mass might also explain baryogenesis through an initial leptogenesis mechanism, and could even be associated with the breaking of such fundamental symmetries as $CPT$ or Lorentz invariance.

There have been many classes of models for neutrino mass (for reviews, see, e.g.,~\cite{Mohapatra:2005wg,GonzalezGarcia:2007ib,Langacker:1226768}), 
including the minimal seesaw~\cite{Minkowski:1977sc,gmrs79,Yanagida:1979as,Schechter:1980gr},  
variations  in which the role of sterile neutrinos is played by neutralinos in $R$-parity violating supersymmetry, extended seesaws involving additional mass scales, and Higgs triplet models. There are also models with small Majorana and/or Dirac masses  generated by higher-dimensional operators or by loop effects (usually with the lowest-order terms forbidden by new symmetries), models of the flavor structure associated with broken family symmetries, and models involving hidden or mirror sectors, large or warped dimensions, or nonperturbative effects.

Especially popular are the minimal seesaw models, in which a heavy Majorana sterile neutrino mixes with an active neutrino.
The heavy scale may be as low as a TeV, but it is  often assumed to be associated with an $SO(10)$ GUT model, perhaps combined with a flavor symmetry. Such models are very attractive, since no unbroken gauge symmetries forbid such sterile Majorana masses, and it is generally believed that ``if they are not forbidden than they are obligatory''.  The minimal seesaw also
connects nicely with leptogenesis~\cite{Fukugita:1986hr}, and it is in some sense a minimal extension of the standard model (SM).

However, the minimal seesaw is  not the only possibility. Some of the assumptions, such as a large Majorana mass, may not be allowed by new symmetries at the TeV scale. The assumption that Majorana masses are obligatory also needs to be examined: in a theory not including gravity they could be forbidden by an unbroken global lepton number ($L$) symmetry, unless some new $L$-violating ingredient (such as a Higgs 126-plet  in $SO(10)$) is added explicitly\footnote{Global symmetries are thought to be violated by gravitational effects~\cite{Banks:1988yz,Witten:2000dt}, but the simplest assumption is that gravity would induce  Majorana masses  $m_\nu \lesssim \nu^2/ \overline M_P \sim 10^{-5}$ eV or smaller~\cite{Weinberg:1980bf},  where $\nu\sim 246$ GeV is the electroweak scale and 
 $\overline{M}_P \sim 2 \x 10^{18}$ GeV
  is the reduced Planck scale. This is much smaller than the observed scale.}.
Moreover, some of the ingredients, 
such as the large Higgs representations that are often invoked in $SO(10)$, or even the existence of a separate stage of grand unification in the effective four-dimensional theory, do not always emerge easily in superstring constructions. While some form of the seesaw may occur for some string vacua, it may be very different in detail from the conventional ``bottom-up'' version. Furthermore, alternatives such as small Dirac masses from higher-dimensional operators, small Majorana {\em or} Dirac masses from exponentially suppressed nonperturbative string instantons, or wave function overlap effects in extra-dimensional theories, have been suggested.  Some of these string-motivated possibilities suggest new
TeV-scale physics and symmetries, which may have implications for the LHC, electroweak baryogenesis, etc.
It is the purpose of this article to explore some of these ideas.

\subsection{Basic Concepts for Neutrino Mass}\label{basics}
A {\em Weyl fermion} is the minimal fermionic degree of freedom. It consists of two components: e.g., a left-chiral fermion $\psi_L$
(satisfying $P_L \psi_L =\psi_L$), and its $CP$-conjugate, the right-chiral antifermion  $ \psi^c_R= P_R \psi^c_R$. (It is  a matter of definition which is called the fermion.)
The chirality
projection $P_{L,R}= \oh (1\mp \gamma^5)$ coincides with helicity for a massless particle. $ \psi^c_R$ is essentially the Hermitian conjugate\footnote{Explicitly, $\psi^c_R=\mathcal{C} \gamma^{0T} \psi_L^\ast$, where $\mathcal{C}$ is the charge conjugation matrix ($\mathcal{C}=i \gamma^2 \gamma^0$ in the Pauli-Dirac representation) and $\psi^\ast_{L\alpha}\equiv\left( \psi_{L\alpha}\right)^\dagger$} of $\psi_L$, and exists whether or not $CP$ invariance holds. 

An {\em active} (also known as  ordinary or doublet) neutrino, is in an \st\ doublet with a charged lepton, so that it has conventional charged and neutral current weak interactions. There are three light active neutrinos $\nu_{aL}, a= e, \mu, \tau$
(the \st\  partners of $e^-_L, \mu^-_L,$ and $\tau^-_L$, respectively), and their $CP$-conjugates $\nu^c_{aR}$, the partners of
$a^+_R$. Any additional active neutrinos would have to be very heavy ($m_\nu \gtrsim M_Z/2$) because of the observed $Z$-width~\cite{Nakamura:2010zzi}.

 A {\em sterile} (a.k.a.  singlet or right-handed)  neutrino is an \st\ singlet. It therefore has no  interactions except for Yukawa couplings to the Higgs, interactions due to mixing, or new interactions beyond the standard model.
A right-chiral sterile $\nu_R$ has a left-chiral $CP$ conjugate $\nu^c_L$. Most extensions of the SM involve sterile neutrinos. Models are often distinguished by whether the sterile neutrinos are light or heavy, and whether they mix with active neutrinos.

A  {\em fermion mass} term in the Lagrangian density describes a transition between right and left Weyl spinors:
\beq -\lag=m \bar{\psi}_L \psi_R+ m^\ast \bar{\psi}_R\psi_L  \ra m  \left(  \bar{\psi}_L \psi_R+  \bar{\psi}_R\psi_L \right), \eeql{fmass}
where $m$ can be taken to be real and non-negative by choosing an appropriate  relative $\psi_{L,R}$ phase. 
An intuitive connection is that a massive fermion that is right-handed in one Lorentz frame is left-handed in another frame in which the three momentum is reversed, while the helicity of a massless particle (which has no rest frame) is invariant.

\begin{figure}
\centerline{\psfig{figure=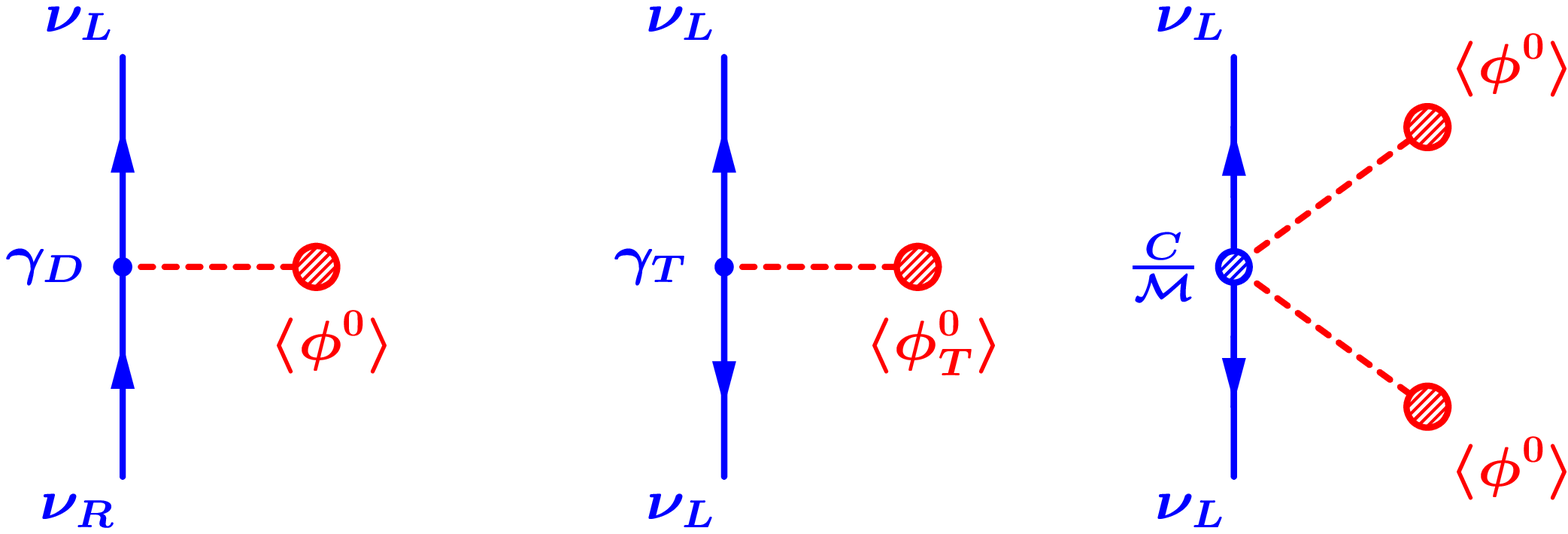,height=11pc}}
\caption{Left: Dirac mass term generated by a Higgs doublet.
Center: Majorana mass term generated by a Higgs triplet. Right: Majorana mass term generated by a higher-dimensional operator.}
\label{fig1}
\end{figure}

A {\em Dirac mass} connects two distinct Weyl spinors (usually one active and one sterile):
\beq -\lag_D= m_D \left( \bar{\nu}_L \nu_R +\bar{\nu}_{R}\nu_{L}\right)= m_D \bar \nu_D \nu_D, \eeql{diracmass}
where, in the second form, $\nu_D \equiv \nu_L+\nu_R$ is the {\em Dirac field}. $\nu_D$ (and its
conjugate $\nu^c_D=\nu^c_R+\nu^c_L$) have four components, $\nu_L, \nu_R, \nu^c_R$, and $\nu^c_L$.  The mass term
conserves lepton number ($L=1$ for  $\nu_D$ and $L=-1$ for   $\nu^c_D$), but violates the third component of weak isospin by $\Delta t^3_L=\pm \oh$ (assuming $\nu_L (\nu_R$) is active (sterile)). It can be generated by the vacuum expectation value (VEV) of a Higgs
 doublet, just as for the quarks and charged leptons, as illustrated in Figure \ref{fig1}, yielding
\beq   m_D =\gamma_D  \veva{\phi^0}\equiv \gamma_D v,  \eeql{Higgsdirac}
where $\gamma_D$ is the Yukawa coupling and $ \rt\, v = \nu \sim 246$ GeV is the weak scale. The problem is understanding
why $\gamma_D$ is so small, e.g.,  $\gamma_D \sim 10^{-12}$ for $m_D \sim 0.1$ eV, as compared to $\gamma_e\sim 3\x  10^{-6}$ for the electron. Possible explanations for a small Dirac mass include higher-dimension operators, string instantons, or large or warped extra dimensions.

A {\em Majorana mass} describes a transition from a Weyl spinor into its own $CP$ conjugate. Majorana mass terms can be written for either active neutrinos ($\nu_L$)
or sterile neutrinos ($\nu_R$):
\beq \begin{split}
-\lag_T & =  \frac{m_{T}}{2}\left(\bar{\nu}_{L}\nu_{R}^{c}+\bar{\nu}_{R}^{c}\nu_{L}\right)
= 
\frac{m_{T}}{2}\left(\bar{\nu}_{L}\mathcal{C}\bar{\nu}_{L}^{T}+\nu_L^T\mathcal{C} \nu_L\right)= 
\frac{m_{T}}{2}\bar{\nu}_M\nu_M \qquad
   \text{(active)} \\
   -\lag_S &=  \frac{m_{S}}{2}\left(\bar{\nu}_{L}^{c} \nu_{R}+\bar{\nu}_{R} \nu_{L}^{c}\right)
   = \frac{m_{S}}{2}\left(\bar{\nu}^c_{L}\mathcal{C}\bar{\nu}_{L}^{cT}+\nu_L^{cT}\mathcal{C} \nu^c_L\right)=\frac{m_{S}}{2}\bar{\nu}_{M_S}\nu_{M_S}
    \quad \text{(sterile)}.
\end{split} \eeql{Majoranamass}
The second forms emphasize that a Majorana mass term also describes the creation or
annihilation of two neutrinos or antineutrinos. In the last forms,
 $\nu_M\equiv\nu_{L}+\nu_{R}^{c}=\nu^c_M$ and 
$\nu_{M_S} \equiv \nu^c_L+\nu_R= \nu^c_{M_S}$
are {\em Majorana fields} for the active and sterile states, respectively.
Each is self-conjugate and has  two-components.  Majorana mass terms violate lepton number by $\Delta L=\pm 2$,
and can therefore lead to neutrinoless double beta decay ($\beta\beta_{0\nu}$).
An active Majorana mass term also violates weak isospin by $\Delta  t^3_L = \pm 1$ (hence the subscript $T$, for triplet), and can be generated by
a Higgs triplet, with a small Yukawa coupling $\gamma_T$ and/or a small VEV \veva{\phi^0_T}, as illustrated in Figure \ref{fig1}.
Alternatively, it can be due to a higher-dimensional operator involving two Higgs doublets, with a coefficient
$C/\mathcal{M}$, where $\mathcal{M}$ is typically associated with a (large) new physics scale.
Majorana mass terms for a sterile neutrino are \st\ invariant. In principle they can be due to a bare mass term, but in most models 
these are forbidden by new physics constraints,  and are instead generated by, e.g.,  the VEV of a SM Higgs singlet field $S$.
Unlike the Dirac case, the phase of a Weyl spinor with a Majorana mass is fixed by the requirement that the
mass is real and positive, leading   to extra observable (in principle)
{\em Majorana phases} in $\beta\beta_{0\nu}$ in the extension with two or more families.

Dirac and Majorana mass terms can be present simultaneously. For one active and one sterile neutrino,
\beq -\lag=\frac{1}{2}
\left(\begin{array}{cc}
\bar{\nu}_{L}^{0} & \bar{\nu}_{L}^{0c}\end{array}\right)
\left(\begin{array}{cc}
m_{T} & m_{D}\\
m_{D} & m_{S}
\end{array}\right)
\left(\begin{array}{c}\nu_{R}^{0c} \\ \nu_{R}^{0}\end{array}\right) + h.c.,
  \eeql{simultaneous}
where the superscripts imply that these are {\em weak eigenstates}. The (symmetric)
mass matrix must be diagonalized, leading, in the general case, to two Majorana mass eigenstates
$ \nu_{iM}=\nu_{iL}+\nu_{iR}^c=\nu_{iM}^{c}, \ i=1,2$,
where 
\beq 
\left(\begin{array}{c}\nu_{1L} \\ \nu_{2L} \end{array}\right)=
A_{L}^{\nu \dag} \left(\begin{array}{c}\nu_{L}^{0} \\ \nu_{L}^{0c}\end{array}\right), \qquad
\left(\begin{array}{c}\nu_{1R}^c \\ \nu_{2R}^c \end{array}\right)=
A_{R}^{\nu \dag} \left(\begin{array}{c}\nu_{R}^{0c} \\ \nu_{R}^{0}\end{array}\right).
\eeq
 $A_L^\nu=A_R^{\nu \ast} $ are unitary matrices, and the mass eigenvalues $m_{1,2}$ are 
related by
\beq  A_{L}^{\nu \dag}
{ \left(\begin{array}{cc}
m_{T} & m_{D}\\
m_{D} & m_{S}
\end{array}\right)}
A_{R}^{\nu}=  \left(\begin{array}{cc}
m_{1} & 0\\
0 & m_{2}
\end{array}\right). \eeql{masseigenvalues}
Important special cases include: (a) Pure Majorana, $m_D=0$. (b) Pure Dirac, $m_T=m_S=0$. One sees from \refl{simultaneous} that a
Dirac neutrino can be thought of as two degenerate Majorana neutrinos (i.e, $m_1=-m_2$ before redefining phases)
with $45^\circ$ mixing. The two mass eigenstate contributions cancel in  $\beta\beta_{0\nu}$, leading to a conserved lepton number. (c) The pseudo-Dirac limit, $m_D \gg m_{T,S}$, which leads to a small shift in the mass eigenvalues
$ |m_{1,2}|=m_D\pm (m_T + m_S)/2$. (We are taking the masses here to be real and positive for simplicity.) (d) The seesaw limit, $m_S \gg m_{D,T}$.  There is one, mainly sterile, state
with $m_2 \sim m_S$, which decouples at low energy (but may play a role in leptogenesis), and one light (mainly active)
state, with mass $m_1 \sim m_T - m_D^2/m_S$. For $m_T=0$, this yields an elegant explanation for why
$|m_1| \ll m_D$. The sterile state may be integrated out, leading to an effective higher-dimensional operator
for the active Majorana mass with scale $\mathcal{ M} \sim m_S$ (more precisely, $\mathcal{ M}/C= \veva{\phi^0}^2 m_S/m_D^2$).

The mass matrix in (\ref{simultaneous}) may be extended to, e.g, three active and three sterile states, leading, in general,
to six Majorana mass eigenstates. In the pure Majorana (with small $m_T$ and large $m_S$), pure Dirac  (with small $m_D$), and seesaw
limits  there are three light active mass eigenstate neutrino fields $\nu_{iL}$ 
that are related to the weak eigenstate fields $\nu_{aL}$ by  $\nu_{aL}= \sum_{i=1}^3 U_{ai}\,  \nu_{iL}$,
where $U$ is the unitary PMNS matrix~\cite{Pontecorvo:1967fh,Maki:1962mu}, and similarly for their  $CP$ conjugates $\nu^c_{iR}$.

\subsection{Top-Down Models}
Superstring theories (for introductions, see, e.g.,~\cite{Becker:1003112,Kiritsis:2003mc,Blumenhagen:2005mu,Blumenhagen:2006ci,Nilles:2008gq,Cvetic:2011vz}) 
are perhaps the most promising and ambitious ultimate extensions of the SM: they are mathematically-consistent finite theories that treat matter and the
 microscopic interactions on a uniform footing, and they naturally incorporate quantum gravity. There are actually several classes of string theories, thought to be different limiting cases of a somewhat mysterious underlying {\em M theory}
and related to each other by {\em dualities}, e.g., one weakly-coupled theory is the strong-coupling limit of another. However, string theories are extremely complicated. Consistency requires   six extra space dimensions, which are presumably compactified on a scale $R \gtrsim \overline{M}_P^{-1}$, where  $\overline{M}_P=M_P/\sqrt{8\pi} \sim 2.4 \x 10^{18}$ GeV is the reduced Planck scale. The {\em string scale} $ M_{s}$ is typically assumed to be comparable  to $\overline{M}_P$, but could be much smaller if some of the extra dimensions are much larger than the others. 
In general, up to constants, one expects
\beq M_s^{2+\delta} \sim g^2_s\, \overline{M}^2_P/V_\delta,
\eeql{couplingdefs} 
 where $V_\delta \sim R^\delta$ is the volume of $\delta$ internal dimensions and $g_s \lesssim 1$ is the  string interaction strength.
Complicated manifolds for the extra dimensions are required to obtain realistic theories (e.g., allowing chiral fermions). These include the large class of {\em Calabi-Yau} manifolds and the somewhat simpler {\em orbifolds}. In the latter, points related by discrete transformations are identified, and string calculations can be performed using conformal field theory techniques.

In recent years, the situation  has become  more difficult
by the realization that there is an enormous {\em landscape} of possible string vacua (perhaps $10^{600}$ or more), differing in part by the topologies, sizes, and shapes of the extra dimensions and by $g_s$. These are determined and parametrized by the VEVs
of scalar {\em moduli} fields, which have no perturbative potential but may be stabilized by nonzero background {\em fluxes} of tensor fields. Most of these vacua do not resemble our world, but a certain fraction do. There has been considerable
effort in exploring particular vacua or classes of vacua that contain the SM or its minimal supersymmetric extension (the MSSM),
with the hope of gaining insight into the SM/MSSM parameters, supersymmetry breaking and mediation, etc. This exploration of the friendly parts of the landscape has so far not produced any completely predictive, compelling, or unique models, but 
some interesting results have nevertheless emerged. For example, vacua containing the MSSM often
contain {\em remnants} (e.g.,~\cite{Blumenhagen:2005mu,Anastasopoulos:2006da,Barger:2007ay,Langacker:2008yv,Cvetic:2011iq}), which are extensions of the MSSM that remain at low scales due to accidents of the
string compactification and do not necessarily solve any problems of the standard model. Especially common are additional
\zpr\ gauge bosons, extended Higgs sectors,  new quasi-chiral exotics  (fermions which occur as vector pairs under the SM, but may be chiral under additional symmetries), and new quasi-hidden sectors, such as those that have been suggested for dark matter or supersymmetry breaking. On the other hand, some SM extensions, such as large representations of gauge groups, are unlikely to occur, at least for the familiar constructions.
Study of the landscape may also suggest new physical mechanisms, such as  nonperturbative string instantons,
``stringy''  higher-dimensional operators, or large extra dimensions, which will be a major focus of this article.
Finally, it should be mentioned that the large number of possible vacua suggests that some physical parameters may be
{\em environmental}, i.e., determined by an accident of the location in the landscape rather than an absolute principle. When combined with some form of eternal inflation~\cite{Linde:1986fc} this may also justify an anthropic approach~\cite{Weinberg:1988cp} to certain problems\footnote{An anthropic explanation for small neutrino masses has been suggested in~\cite{Pogosian:2004hd}.}, such as the magnitude of the dark energy. 

The ``oddball'' neutrino masses are especially interesting in this connection since they may be directly connected to high-scale physics and since there are so many possibilities depending on the point in the string landscape. In principle, a given string vacuum would uniquely predict all details of the neutrino spectrum, such as the type of hierarchy, mixing angles, Majorana or Dirac nature, and overall mass scale. This is to be contrasted with grand unification and most bottom-up models, which {\em by themselves} shed little light on the details of the hierarchy and mixings until supplemented with extra family symmetries and other model-dependent assumptions. In practice, however, 
the landscape is so varied that it is difficult to construct fully realistic models or to make reliable predictions about the details of the neutrino (or other particle) spectrum. We will therefore concentrate  instead on general mechanisms for small neutrinos masses, and related issues such as the underlying scales, the role of $L$ conservation or violation, and the possible connection with new TeV-scale physics and symmetries.

\subsection{Superstrings}\label{stringdescription}
\begin{figure}
\centerline{\psfig{figure=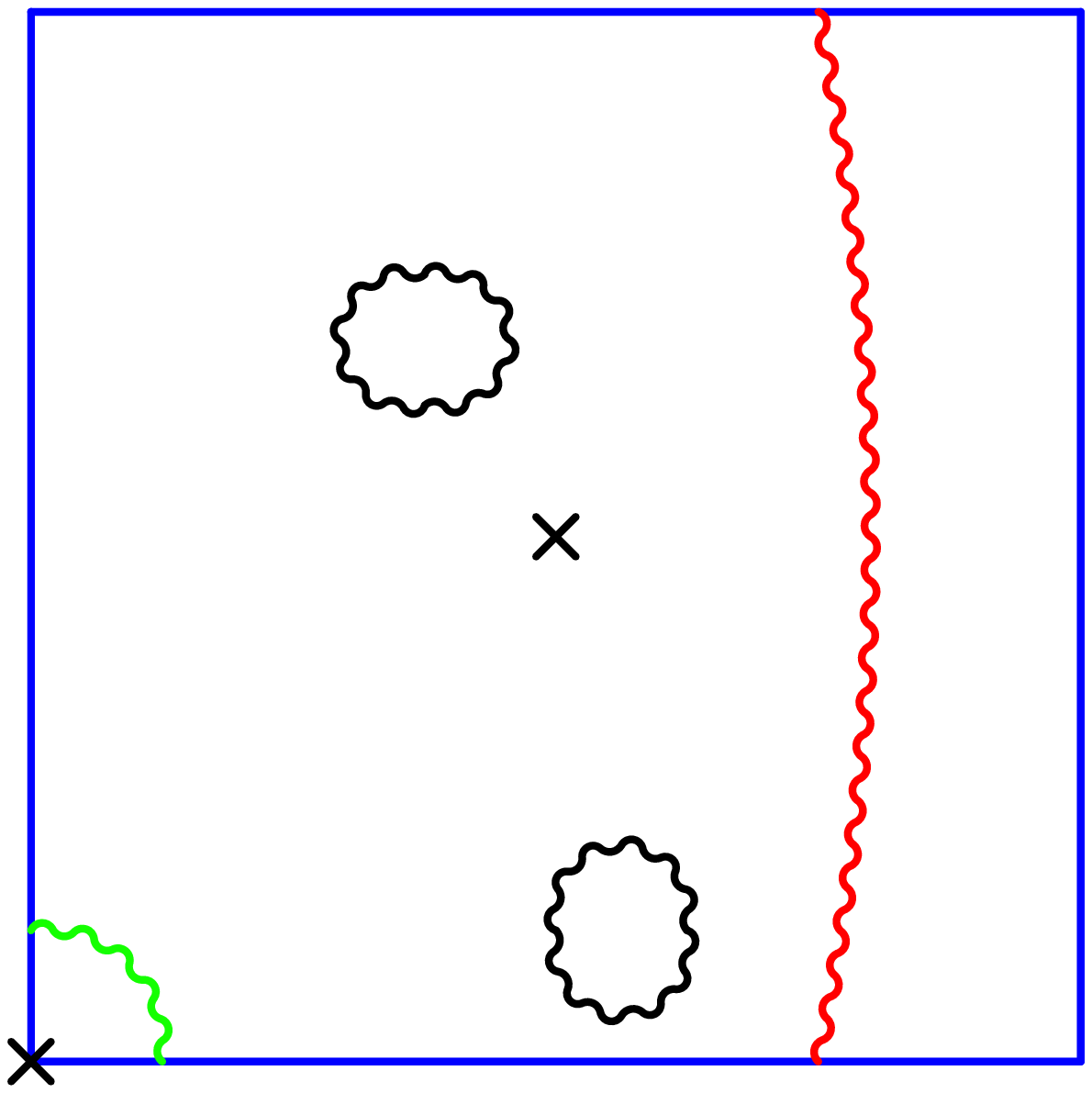,height=12pc}\hspace*{1cm}
\psfig{figure=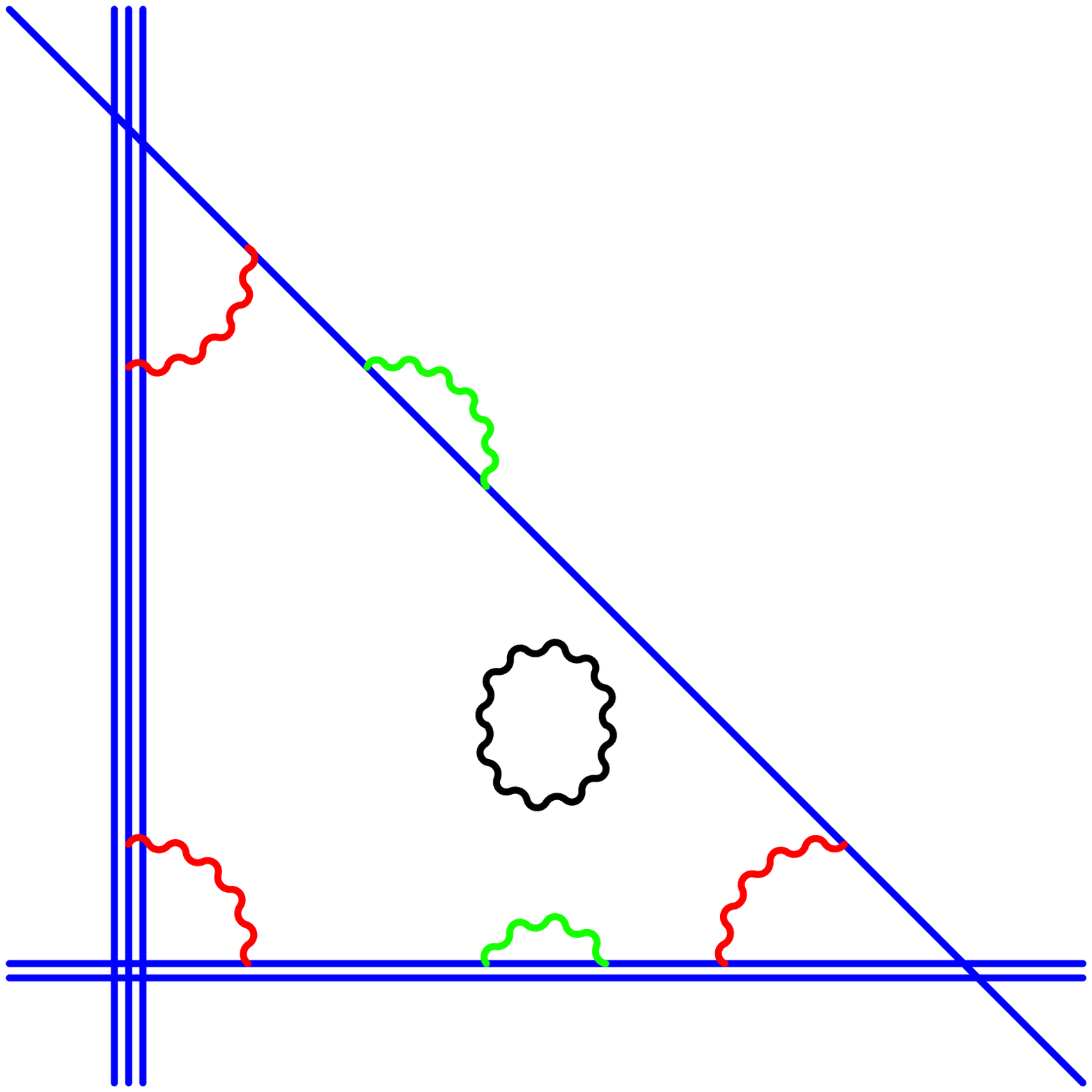,height=12pc}}
\caption{Left: An unwrapped torus, representing two compactified dimensions. The bottom and left edges are related by an additional orbifold symmetry (rotation by $\pi/2$), with fixed points at the corners and center.
Shown are  freely propagating closed strings  (black), a winding mode (red), and a twisted state (green). Right: Stacks of intersecting D-branes, with gauge bosons (green) terminating on a stack,
matter fields (red) localized at intersections, and gravitons (black) propagating freely.}
\label{combinedstrings}
\end{figure}

Strings may be thought of as one-dimensional vibrating objects, which may be open or closed. Ordinary particles, gauge bosons, and gravitons, which appear to be point-like on scales large compared to $M_{s}^{-1}$,
correspond to massless vibrational modes, while string excitations are modes with mass of  \ord{M_{s}}.
The {\em heterotic} string theories~\cite{Gross:1984dd}, based on $E_8\x E_8$ or $SO(32)$, involve closed strings. These  states may propagate freely in the extra dimensions with zero momentum or have non-zero quantized momentum
of \ord{j /R} ({\em Kaluza-Klein excitations}). There are also {\em winding} modes of energy \ord{k R\, T},
where $ T \sim M_s^{2}$ is the string tension, corresponding to the string winding $k$ times around the compact dimension.
We also mention
  {\em twisted} modes, in which the string is trapped at a fixed point of, e.g., the orbifold point transformation.
 These are illustrated in Figure \ref{combinedstrings}.
 
Another important class are  
 Types I, IIA, and IIB theories. These involve both closed strings (for gravitons) and open strings which start and end on nonperturbative extended objects known as D$p$ {\em branes}, which fill $p$ space dimensions. For example,
 in the type IIA intersecting brane constructions (see   Figure \ref{combinedstrings}) there are stacks of $N$ parallel D6 branes (which fill the ordinary and 3 of the 6 extra space dimensions), corresponding to the gauge groups $U(N)$. Gauge bosons are described by strings terminating on branes within the same stack, while matter fields are localized at the intersections of two stacks.
 The latter are  bifundamental, such as $(N,M)$ or $(\overline N,M)$ for $U(N)$ and $U(M)$ stacks. Families of particles are associated with multiple intersections of the stacks  as they wind around the extra dimensions.
 More complicated {\em orientifold}  compactifications can also
 lead to symmetric or antisymmetric fields under $U(N)$, allowing for example the antisymmetric 10 representation of the $U(5)$ (grand unification) group, or to  $Sp(2 N)$ or $SO(2N)$ gauge groups. 


\subsection{Neutrino Mass Suppression Mechanisms}\label{suppression}
Consider a generic Yukawa coupling  and fermion mass term for Weyl fermions $\psi_{L,R}$ and complex scalar $\varphi$:
\beq
\lag=\gamma \bar \psi_L \psi_R \varphi -m \bar \psi_L \psi_R  + h.c.
\eeql{generic}
We wish to consider plausible mechanisms in which $\gamma$ and/or  $m$ are extremely small, e.g.,
$\gamma= \ord{10^{-12}}$ and $m =\ord{\text{eV}}$.
One can of course simply choose these as the renormalized values of the parameters,
and in fact (in this example) they are protected against large radiative corrections because  the symmetry of the theory increases when they
vanish (i.e., they are  {\em technically natural}~\cite{tHooft:1979bh}). However, the small values may instead parametrize some underlying mechanism.

One possibility is that the terms in \refl{generic} are  forbidden by some additional beyond the standard model symmetry (gauge, global, or discrete), but that  {\em higher-dimensional operators} (HDO) such as
\beq
\lag=c_p \frac{S_1 S_2\cdots S_p}{\mathcal{M}^p} \bar \psi_L \psi_R \varphi - d_q\frac{S'_1 S'_2\cdots S'_{q+1}}{\mathcal{M}^q}  \bar \psi_L \psi_R  + h.c.,
\eeql{hdo}
are allowed\footnote{Such HDO are often invoked in the  Froggatt-Nielsen mechanism~\cite{Froggatt:1978nt} for generating fermion mass hierarchies via a broken family symmetry. Similarly, the chiral \sto\ symmetry of the SM forbids bare fermion masses, which are instead generated by introducing the Yukawa coupling to the Higgs doublet $\varphi$, analogous to replacing $S$ by $\varphi$ in the second term, with $q=0$.}.
Here, $S_i$ and $S'_j$  are SM singlet fields that are charged under the new symmetry,  $\mathcal{M}$ represents some  underlying new physics scale that has been integrated out, and $c_p$ and $ d_q$ are dimensionless coefficients, which are sometimes absorbed into \mcal. (For simplicity we will denote the operators in \refl{hdo} by $S^p/\mathcal{M}^p$ and $S^{q+1}/\mathcal{M}^q$.)
For example, the effective Yukawa for $p=1$ can be generated by mixing $\psi$ with a heavy fermion $\Psi$, with
\beq \lag=\gamma_\varphi \bar\psi_L \Psi_R \varphi + \gamma_S \bar\Psi_L \psi_R S - M_\Psi \bar\Psi_L \Psi_R + h.c.,
\eeql{heavyPsi}
implying $\mathcal{M}=M_\Psi$ and $c_1=\gamma_\varphi \gamma_S$. 
When $S$ in \refl{hdo} acquires a VEV the
symmetry is broken, yielding an effective Yukawa coupling $c_p \veva{S}^p/\mathcal{M}^p$ or mass $d_p \veva{S}^{q+1}/\mathcal{M}^q$,
which can be very small for $\veva{S} \ll \mathcal{M}$ and $p \ge 1$  or $q\ge 1$.

The HDO may be associated with heavy states that are exchanged at tree level or which occur in loops. If the state is a particle that is part of the underlying four-dimensional field theory (as in the toy example above or in GUT models) it will be referred to as {\em field theoretic}.
If it is associated directly with an underlying string  theory (e.g., by the exchange of string excitations) it will be called {\em stringy},
with $\mcal\sim M_{s}$ or \mred.
In heterotic models the hierarchies of, e.g., quark and charged lepton masses are often associated with stringy HDO,
 typically with $\veva{S}/M_{s} \sim 1/10$ for some of the $S$ fields.
 
Another possibility for small couplings or masses is due to geometric suppressions associated with extra dimensions\footnote{The coefficients of the HDO in heterotic theories are also associated with the extra dimensions, but here we refer to more manifestly geometric considerations.}. Interaction strengths in theories with large or warped extra dimensions are proportional to the wave function overlaps in those dimensions, which can be very small, e.g., if some of the particles
are confined to the boundary of a large dimension (the {\em brane}), while others  can propagate freely
in the {\em bulk} of the dimension. Similarly, in the intersecting brane constructions, when three stacks of D6 branes form a triangle in the extra dimensions, as in Figure \ref{combinedstrings}, there will be a Yukawa interaction between the
states at the corners. This is related to the action of the string worldsheet stretching between the corners, and is
proportional to $e^{-A}$, where $A$ is the area of the triangle. Hierarchies of interaction strengths are therefore associated
with geometric area factors, and very small  couplings can be obtained for large $A$. More complex nonperturbative {\em D instanton} suppressions will be introduced in Section \ref{Majorana}.

\section{SMALL MAJORANA MASSES}\label{Majorana}

Many models for Majorana neutrino masses can be parametrized by the Weinberg operator~\cite{Weinberg:1980bf}
\beq  -\lag_{\phi\phi} =\frac{C}{2\mcal} \left(  \bar \ell_L  \vec{\tau} \,  \tilde \ell_R \right)  \cdot \left( \phi^\dag \vec\tau \tilde \phi \right) +h.c. 
= \frac{C}{\mcal} \left(  \bar \ell_L   \,  \tilde \phi\right)  \left( \phi^\dag \,  \tilde \ell_R  \right) +h.c. \eeql{Weinbergop}
with  two Higgs doublets  arranged to transform as an \st\ triplet. In \refl{Weinbergop},
\beq \ell_L\equiv \douba{{\nu}_L}{{e}_L}, \quad \tilde \ell_R \equiv  \douba{e^{c}_R}{-\nu_{R}^{c}}, \quad \phi \equiv\douba{\phi^+}{\phi^0}, \quad \tilde \phi\equiv\douba{\phi^{0\dag}}{- \phi^-}, \eeql{fielddef}
and lepton flavor indices are suppressed.
In supersymmetric notation, the corresponding superpotential term is 
\beq W= \frac{C}{\mcal}\, L H_u\,  L H_u, \eeql{Weinbergsuper}
where $L\equiv (N \ E)^T$ and $H_u\equiv (H^+_u\ H^0_u)^T$ are respectively the lepton and up-Higgs chiral supermultiplets.

\subsection{The GUT Seesaw}
\begin{figure}
\centerline{\psfig{figure=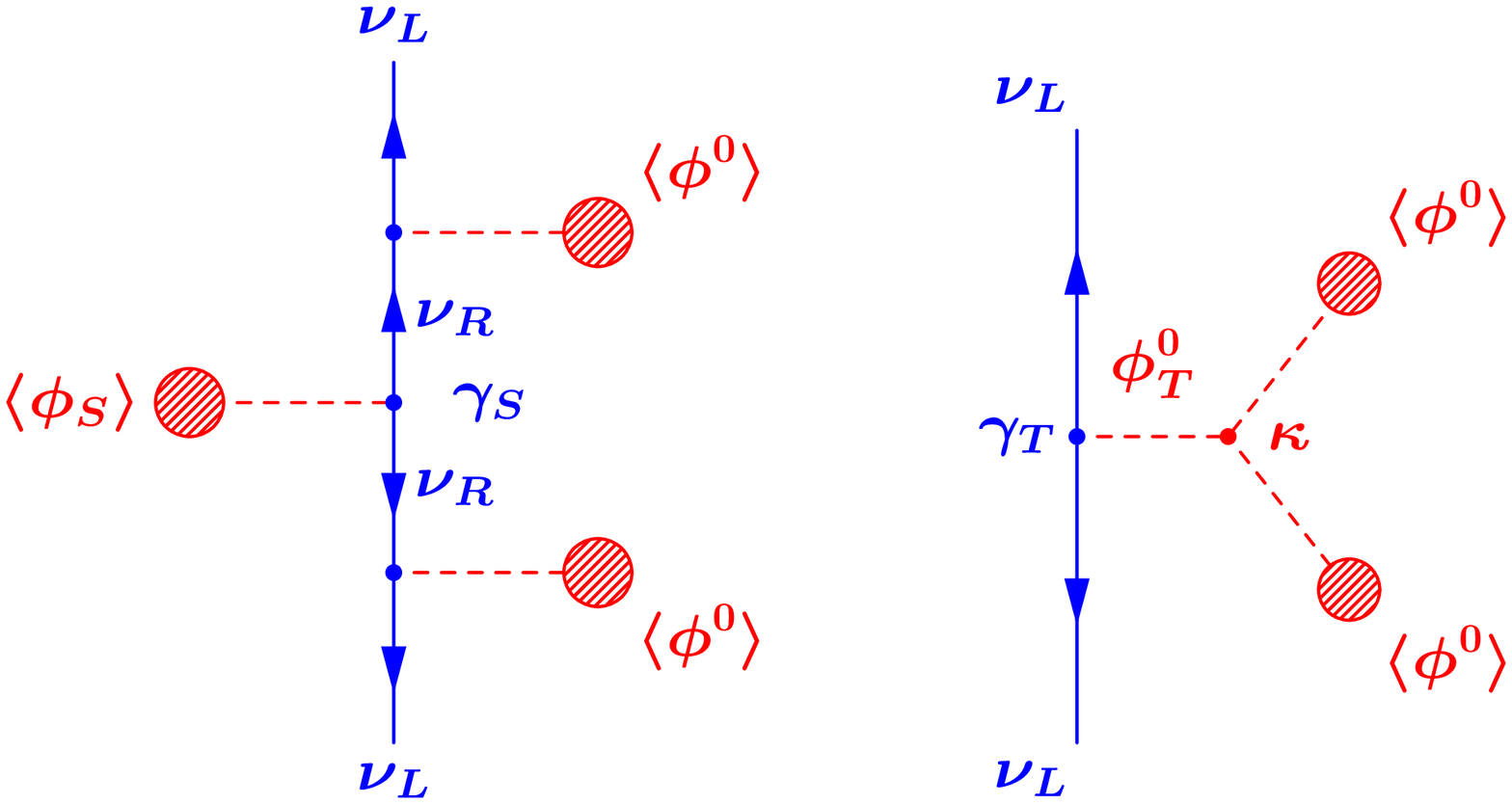,height=17pc}}
\caption{Type I  (heavy sterile $\nu_R$) and type II (heavy scalar triplet) seesaws.
$\phi_S, \phi,$ and $\phi_T$ refer respectively to singlet, doublet, and triplet Higgs fields,
while $\gamma_{S}, \gamma_T,$ and $\kappa$ are couplings.
}
\label{gutseesaws}
\end{figure}

The  Weinberg operator may be generated by heavy particle exchange in  the effective four-dimensional theory,
such as in grand unification~\cite{Mohapatra:2005wg,GonzalezGarcia:2007ib,Langacker:1226768,Raby:2008gh}. For example, $SO(10)$ GUTs can lead to either the type I (heavy Majorana sterile neutrino)
or type II  (heavy triplet) seesaws (or both), as illustrated in Figure \ref{gutseesaws}.


In $SO(10)$ each family of fermions (including $\nu_R$) is assigned to a ${16}-$plet, while the Higgs doublets are contained in a ${10}$-plet, $10_H$. The simplest way to generate  a large Majorana mass for $\nu_R$ at the renormalizable level is to introduce  a 126-dimensional Higgs representation ${126}^\ast_H$, which contains a SM singlet $\phi_S$
that is assumed to obtain a VEV near the GUT scale, i.e.,
\beq  W\sim \gamma_u \, \underbrace{16 \x 16 \x 10_H}_{\bar \nu_L \nu_R \phi^0}
+ \gamma_S \,  \underbrace{16 \x 16 \x 126_H^\ast}_{\bar \nu^c_L \nu_R \phi_S},
\eeql{so10}
yielding a sterile Majorana mass $m_S /2\sim \gamma_S \veva{\phi_S}$.
An alternative to the ${126^\ast_H}$-plet~\cite{Babu:1998wi} is to introduce  an additional Higgs $16^\ast_H$ with a large VEV for the SM-singlet component,
and assume a HDO
\beq
W\sim \gamma_u \, 16 \x 16 \x 10_H+
\frac{c}{\mcal}\,  16 \x 16 \x 16^\ast_H\x 16^\ast_H.
\eeql{guthdo}
This could be associated with a hybrid model, in which the HDO in \refl{guthdo} is stringy, while the Weinberg operator
obtained by integrating out the $\nu_R$ is field-theoretic. Alternatively,  \refl{guthdo} could derive from integrating out a still
heavier $SO(10)$ singlet~\cite{Kim:2003vr}, i.e.,  a two-stage field theoretic model.
One requires 
$\gamma_S|\langle \phi_{S}\rangle |$ or ${c}\, |\langle \phi_{16^\ast_H}\rangle|^2/{\mcal}
\sim 10^{14} \text{ GeV}\ |\gamma_u \langle \phi^0\rangle/{100\text{ GeV}}|^2$
for an active neutrino mass $\sim 0.1$ eV.
Neither of these scenarios has been implemented in a full string construction.

Orbifold GUTs are reviewed in~\cite{Raby:2008gh,Kim:2003vr}.
 
\subsection{The Heterotic Seesaw}\label{heterotic}
String theories often involve an underlying grand unification. However, a full GUT theory does not necessarily survive into the effective four-dimensional field theory. Even if the GUT gauge symmetry is present in a construction, the Higgs representations needed to break the GUT group or to generate large Majorana neutrino masses by a type I seesaw may be absent. For example,
the $E_8 \x E_8$ heterotic string may lead to an $E_6$ GUT (or its $SO(10)$ subgroup) in the effective theory~\cite{Dine:1985vv,Breit:1985ud,Witten:1985bz,Cecotti:1985by}. 
However, the simplest expectation is that the matter and
Higgs fields are contained in $27$-plets. While one can obtain small or vanishing neutrino masses by fine-tuning~\cite{Dine:1985vv},  there are no renormalizable couplings contained in $27^3$ (or in $16 ^2 \x 10$ of $SO(10)$) that could lead to large Majorana masses for the $\nu_R$ candidates in the $27$ (or $16$). 
Higher-dimensional Higgs representations, which could in principle
yield Majorana mass terms, are unlikely to emerge from from string theory. For example, the $126$-plets introduced phenomenologically in \refl{so10} are not allowed in known compactifications~\cite{Dienes:1996yh}. 
Because of these difficulties, early $E_6$ string-inspired models often invoked additional fields not in the 
27, such as $E_6$ singlets~\cite{Witten:1985bz,Mohapatra:1986aw,Mohapatra:1986bd},  or higher-dimensional
operators~\cite{Nandi:1985uh}, typically leading to extended versions
of the seesaw model involving fields with masses/VEVs at the TeV scale.

Perhaps more likely
is that the underlying string theory breaks directly to an effective
four-dimensional theory that includes the SM. Some GUT features, especially in the gauge sector, may be maintained. However, the GUT relations for Yukawa couplings are often
not retained~\cite{Dine:1985vv,Breit:1985ud,Witten:1985bz,Cecotti:1985by}.
There may also be stringy constraints on the matter content and Yukawa relations,
 and higher-dimensional
operators,  similar to those in \refl{guthdo}, may occur.

In  supersymmetric constructions, sterile Majorana masses may be generated by
HDO superpotential  terms such as
\beq  W_S = \oh c_S \frac{S^{q+1}}{M_s^q} N^c N^c \Ra m_S=c_S \frac{\veva{S}^{q+1}}{M_s^q},\qquad q\ge 0,
\eeql{Wnur}
where $N^c$ is a left-chiral sterile antineutrino superfield, $S$ is a SM-singlet superfield, $\veva{S}$ is the VEV of its scalar component,  and $c_S$ is a dimensionless coefficient. \refl{Wnur} is easily generalized to a product of $q+1$ distinct
singlet fields, and to include neutrino familiy indices. Similarly, a Dirac mass term is associated with
\beq W_D=c_D  \frac{S^{p}}{M_s^p} L H_u N^c \Ra m_D=c_D \frac{\veva{S}^{p}}{M_s^p} v_u,\qquad p\ge 0,
\eeql{Diracmass}
where $v_u = \veva{H^0_u}$.
A realistic seesaw can occur if the needed operators actually occur in the construction (with appropriate lowest values of $p$ and $q$ allowed by the symmetries of the effective theory and by the string selection rules),  and if the singlet fields 
acquire the necessary VEVs, presumably without breaking supersymmetry or generating a cosmological constant at a
large scale.  One difficulty is to make  $m_S$ sufficiently small, especially for \veva{S} and $M_s$ close to the Planck scale,
e.g., $m_S \sim 10^{14}$ GeV for $p=0$ and $c_D=1$.
Possible mechanisms 
involve approximately flat directions of the potential leading to a small $\veva{S}/M_s$, e.g.,\ associated with
an additional $U(1)'$ gauge symmetry~\cite{Cleaver:1997nj,Langacker:1998ut}; stringy
threshold corrections~\cite{Cvetic:1992ct}; multiple scales~\cite{Haba:1993yj,Haba:1994ym}; or
hidden sector condensates~\cite{Faraggi:1993zh,Coriano:2003ui}.

However, a common situation in heterotic constructions is that the potential is minimized for $\veva{S}/M_s=\ord{1/10}$. This may occur, for example, because there are {\em flat directions}, along which the supersymmetry breaking $F=- (d W/d S)^\ast$
terms and the $D$ terms associated with the extra non-anomalous $U(1)'$ gauge factors (that are common in the constructions) both vanish. Many of these directions also have $W=0$, therefore avoiding a very large ($M_s$-scale) contribution to the
cosmological constant. There is typically one anomalous $U(1)'$ factor with a Fayet-Iliopoulos contribution~\cite{Fayet:1974jb} to its $D$ term.
This forces some of the $\veva{S}$ to be nonzero along the (otherwise) flat directions,
breaking some of the extra gauge symmetries and generating masses for some other particles (i.e., {\em  vacuum restabilization}, e.g.,~\cite{Cleaver:1997jb}). Another possibility is for the $F$ terms to vanish at isolated points, although then the vanishing of $W$ is not so obvious.

There have been a number of studies of the HDO 
such as \refl{Wnur} and \refl{Diracmass}
 that are allowed to various orders by the string constraints and symmetries in several kinds of heterotic constructions, usually imposing
vanishing of the $F$ and $D$ terms, and sometimes other constraints such as $R$-parity conservation.
These include $Z_3$ orbifolds~\cite{Font:1989aj,Giedt:2005vx}, free fermionic models~\cite{Faraggi:1993zh,Coriano:2003ui,Ellis:1997ni,Ellis:1998nk},  a $Z_3$ orbifold with an intermediate
$SU(3)^3$ (trinification)~\cite{Kim:2004pe}, and $Z_3\x Z_2$ orbifolds~\cite{Kobayashi:2004ya,Lebedev:2006kn,Buchmuller:2007zd,Lebedev:2007hv}.
Many  of these studies argued that  Dirac and Majorana mass terms
for a seesaw were possible, often after making additional assumptions. 
However, a detailed study of the $F$ and $D$ flatness conditions in a large class of vacua of the bosonic $Z_3$ orbifold,
including superpotential terms through degree 9, found that only two out of twenty patterns of vacua had Majorana
mass operators, while none had simultaneous Dirac operators of low enough degree to allow neutrino masses larger than
$10^{-5}$ eV. On the other hand, it was shown in~\cite{Lebedev:2006kn,Buchmuller:2007zd} that there is a ``fertile patch''
of the landscape of the  $Z_3\x Z_2$ orbifold in which there are many models involving operators of high dimension
(e.g., $p\sim 1 \cdots 5, \ q \sim 0 \cdots 7$) that can lead to a seesaw. These typically involve mixing with many (\ord{100})
singlet $N^c$ candidates, which may help to increase the mass of the light neutrinos. These often have $F$-flat points and do not all have $W=0,$ but in~\cite{Lebedev:2007hv} two examples with $W=0$ were exhibited.

There are therefore points in the heterotic landscape that lead to a type I seesaw, though how common is still not clear. The known examples differ in detail from GUT-type constructions. In particular, they are hybrid theories typically involving stringy HDOs of rather high degree to generate the Dirac and Majorana masses, which then produce a field theoretic Weinberg operator. Furthermore, they may involve
mixing with a large number of sterile neutrinos, and they do not satisfy simple $SO(10)$ relations.

\subsection{The Seesaw from String Instantons}\label{seesawinstanton}
It is difficult to obtain canonical Majorana mass terms at the perturbative level in Type I or II brane constructions.
As mentioned in section \ref{stringdescription}, gauge symmetries are usually associated with $U(N)$ rather than $SU(N)$.
The extra $U(1)$ factors are  typically anomalous, with the anomalies cancelled by Chern-Simons terms.
(One or more linear combinations, such as weak hypercharge, may be non-anomalous and survive as gauge factors in the low energy theory.) The gauge bosons acquire string-scale masses by the St\"uckelberg mechanism (e.g.,~\cite{Blumenhagen:2005mu}),
but the $U(1)$ symmetries  survive as global selection rules on the perturbative couplings, typically leading to a conserved lepton number\footnote{Heuristically, in the simplest  intersecting brane models one expects superpotential terms to involve three {\em distinct} superfields associated with the corners of a triangle.}. This was shown in detail in two
studies \cite{Ibanez:2001nd,Antoniadis:2002qm}
of nonsupersymmetric
models with a low string scale,
but  holds more generally~\cite{Blumenhagen:2005mu}.

The difficulty of obtaining Majorana masses and other needed couplings was elegantly resolved by
the consideration of nonperturbative string effects.
These are analogous to field-theory instantons (e.g.,~\cite{tHooft:1976fv}), which are exponentially-suppressed
interaction terms associated with localized, topologically non-trivial, classical field configurations. These are typically
of $\ord{e^{-1/g^2}}$, where $g$ is a gauge coupling, and can lead to such effects as $B+L$ violation in the SM.
Similarly, in string theory one encounters worldsheet instantons associated with topologically-nontrivial Euclidean
string configurations localized in space-time~\cite{Dine:1986zy}. D instantons are analogous configurations involving Euclidean D-branes  (for a review, see~\cite{Blumenhagen:2009qh}).
These D instantons may generate perturbatively absent couplings, such as Majorana neutrino masses,
the $\mu$ parameter, or
certain Yukawa couplings~\cite{Blumenhagen:2006xt,Ibanez:2006da,Florea:2006si,Haack:2006cy} 
(see also~\cite{Blumenhagen:2009qh,Cvetic:2007sj,Ibanez:2007rs,Akerblom:2007nh,Blumenhagen:2007zk,Cvetic:2009yh,Kiritsis:2009sf}). 
For example, a Euclidean D2 instanton in a type IIA construction can lead to
couplings that  are exponentially suppressed by
$\exp\, (-S_{inst})$, where  the instanton action $S_{inst}\sim g^{-2}_{YM} V_{E2}/V_{D6}$ is a function of 
gauge coupling $g_{YM}$ 
and the ratio of the volumes of 
 the D2 instanton ($V_{E2}$) and of the relevant D6 branes ($V_{D6}$).  
This is a new stringy effect, which, in contrast to field theory instantons,  allows for new hierarchies of couplings.
In particular, one can obtain a Majorana sterile neutrino mass
\beq m_S \sim M_s e^{-S_{inst}}. \eeql{sterileinstanton}
 For appropriate wrappings of the instanton and brane around the
internal dimensions one can find a value of  $M_S$ that is large but  still sufficiently below the string scale, leading to a type I seesaw~\cite{Blumenhagen:2006xt,Ibanez:2006da,Ibanez:2007rs,Cvetic:2007ku,Antusch:2007jd,Cvetic:2007qj,Ibanez:2008my}. In principle, there could also be a much larger suppression in other constructions, leading to  eV-scale masses (section \ref{smallmixed}). Other applications to neutrino mass are described in sections \ref{nonur} and \ref{Dirac}.

\subsection{Theories with  No  $\nu_R$ or Tree-Level $L N^c H_u$}\label{nonur}
We have seen in sections \ref{heterotic} and \ref{seesawinstanton} that examples of the type I seesaw in string constructions
are typically hybrid, i.e,  combining stringy HDO or D instantons for the large Majorana masses with a field theoretic seesaw operator. Furthermore, such examples differ greatly in detail from typical GUT seesaw models. It is therefore interesting to consider stringy mechanisms that lead more directly to small Dirac or Majorana masses, without the need to introduce a heavy Majorana
sterile neutrino in the effective four-dimensional field theory (or in which sterile superfields $N^c$ are present but
the Dirac couplings are too small  to produce a significant seesaw contribution to the active neutrino mass).

The simplest possibility is the direct generation of the Weinberg operator in \refl{Weinbergsuper}
by stringy effects, such as by the exchange of string excitations, Kaluza-Klein modes, winding modes, moduli, etc., suggesting $\mathcal{M} \sim M_s$. For  $M_s \sim \overline M_P\sim 2.4\x 10^{18}$ GeV and $C \lesssim 1$ the
 resulting  mass is too small,
 $m_\nu \le 10^{-5}$ eV.
This can be remedied by compactifying on a large internal volume $ V_6 \gg M_s^{-6}$,
implying a low string scale 
$M_s \ll \overline M_P$ from \refl{couplingdefs}. For example, it was found in~\cite{Conlon:2007zza}
that $m_\nu \sim 0.1$ eV (i.e., $M_s/C \sim 10^{14}$ GeV) can be obtained in a type IIB flux compactification
for $M_s \sim 10^{11}$ GeV. Although small, this is still much larger than the low  (\ord{\text{TeV}}) scales
often considered in non-supersymmetric constructions~\cite{Ibanez:2001nd,Antoniadis:2002qm} or  large-dimensional scenarios (section \ref{large}).

A stringy Weinberg operator can also be generated by D instantons~\cite{Ibanez:2007rs,Antusch:2007jd,Cvetic:2010mm}. In this case the coefficient is exponentially
suppressed, $C/\mathcal{M} \sim \exp{(-S_{inst})}/M_s$, requiring a low $M_s$.  An analysis of models with 4 and 5 stacks~\cite{Cvetic:2010mm}
of branes showed that the necessary D instantons could be correlated with those generating a $\mu$ term
for $M_s\sim (10^3-10^7)$ GeV, or with  perturbatively-forbidden Yukawa couplings for $M_s\sim (10^9-10^{14})$ GeV.

Several authors have considered neutrino masses in F-theory~\cite{Vafa:1996xn,Heckman:2010bq}, which is a geometric formulation of
strongly coupled type IIB theory in which matter is  localized along  complex curves at the intersections of D7 branes. A stringy Weinberg operator may be generated by  exchange of  complex structure moduli~\cite{Tatar:2009jk}, self-intersecting matter curves~\cite{Randall:2009dw},  D instantons~\cite{Randall:2009dw,Heckman:2008es,Donagi:2011dv}, or Kaluza-Klein modes~\cite{Bouchard:2009bu}. In the latter case, the seesaw mass may be enhanced by the exchange of an infinite tower of modes (cf.~\cite{Lebedev:2006kn}).

Another possibility is to consider a type II seesaw, i.e., involving a heavy Higgs triplet~\cite{Ma:1998dx,Hambye:2000ui}.
It is non-trivial to generate the necessary \st-triplet with non-zero hypercharge in string constructions. In a heterotic construction
one needs  a higher-level embedding of \st, e.g.,  into $\st\x\st$ which is broken to the diagonal subgroup~\cite{Langacker:2005pf}. This leads in lowest order to an
off-diagonal triplet Majorana mass matrix, exhibiting an $L_e-L_\mu-L_\tau$ symmetry in the simplest case. 
This is similar to a well-known bottom-up construction (e.g.,~\cite{Mohapatra:1999zr}), but allows more freedom
in the charged lepton sector. It may also be possible to generate the appropriate triplets in intersecting D6 brane models
involving $U(5)$ grand unification, but the existing examples are not very realistic~\cite{Cvetic:2006by}.

\subsection{Other Higher-Dimensional Operators}\label{otherhdo}
 Majorana ($m_{T,S}$) or Dirac ($m_D$) mass terms may also be generated by other types of HDO.
For example, sterile Majorana masses, feeding into a seesaw, could be generated by operators similar
to \refl{Wnur}, but with \veva{S}\ occurring at a low or intermediate scale. This could occur, for example, if the $q=0$ and bare terms 
(and perhaps the $p=0$ Dirac term) are forbidden
by a (non-anomalous) \uprm\ gauge symmetry broken at an intermediate scale along an approximately flat direction~\cite{Cleaver:1997nj,Langacker:1998ut,Kang:2004ix}. 

A number of authors~\cite{ArkaniHamed:2000bq,Murayama:2004me,ArkaniHamed:2000kj,Borzumati:2000mc,Borzumati:2000ya} have suggested  that small $\nu$ masses may be associated with supersymmetry breaking. Let us parametrize the breaking by a spurion field $X$ with
$\veva{X}=m_X + \theta^2 F_X$ (e.g.,~\cite{Martin:1999hc}). Typically, one expects $F_X/\mathcal{M} \sim m_{soft}$, where $\mathcal{M}$ is the messenger scale and  $ m_{soft}\sim $ TeV is the effective supersymmetry-breaking scale. A  TeV-scale seesaw may be generated in a model~\cite{ArkaniHamed:2000bq,Murayama:2004me} involving operators
\beq
K_S=\frac{1}{\mathcal{M}} X^\dagger N^c N^c+ h.c., \qquad W_D =\frac{1}{\mathcal{M}} X L H_u N^c
\eeql{tevseesaw}
in the K\"ahler potential and superpotential. 
Assuming as well that $m_X^2\sim F_X$ one obtains $m_S\sim  m_{soft}$ and $m_D \sim ( m_{soft}/\mathcal{M})^{1/2}\, v_u$,
leading to a seesaw mass $v_u^2/\mathcal{M}$. For supergravity mediation, one expects
$\mathcal{M}\sim \overline{M}_P$ so that $m_\nu$ is expected to be too small, similar to the stringy Weinberg operator. $m_\nu$ could  be enhanced by a lower $\mathcal{M}$ or other mechanisms~\cite{ArkaniHamed:2000kj}.  Similar considerations can lead to $L$-violating masses for the scalar neutrinos, $\tilde{\nu}_L$,
which can generate $\nu_L$ masses at loop-level by neutralino exchange~\cite{Borzumati:2000mc,MarchRussell:2004uf}.
Non-holomorphic operators, e.g., involving the ``wrong'' Higgs doublet, typically lead to strong suppressions of $\ord{m_{soft}/\mathcal{M}}$~\cite{Martin:1999hc,Borzumati:1999sp}.
This has been exploited in a  model~\cite{Casas:2002sn,Brignole:2010nh} involving the non-holomorphic
K\"ahler term
\beq  K_T \sim \frac{1}{\mathcal{M}^2} LH_u\, L \tilde H_d+h.c., \qquad  \tilde H_d\equiv \douba{H_d^+}{-H_d^{0\dag}},
\eeql{nonhol1}
where $H_d$ is  the down-type Higgs doublet and $\tilde{H}_d$ is essentially its adjoint.
The $F$ component of $H^0_d$ is $F^\ast_{H^0_d}=-\mu H^0_u$, where $\mu= \ord{100\text{ GeV}}$ is
the MSSM $\mu$ parameter. This implies  a suppressed mass $m_\nu \sim {\mu v^2_u}/{\mathcal{M}^2}$, suggesting a mediation scale $\mathcal{M} \sim 10^8$ GeV. 

Other higher-dimensional operators have also been suggested, including those involving quark fields,
which can lead to $\nu$ masses by quark condensation~\cite{Davoudiasl:2005ai} or by loops~\cite{Babu:2001ex,deGouvea:2007xp}, and those leading to TeV-scale seesaws~\cite{Chen:2009fx}. It is also straightforward to construct operators
leading to eV-scale sterile neutrino masses (section \ref{smallmixed}).

\section{SMALL DIRAC MASSES}\label{Dirac}
String constructions, such as those based on intersecting branes, may involve a conserved lepton number that forbids
Majorana masses, at least at the perturbative level. It is also possible that Majorana masses are present but are negligibly small.
It is therefore interesting to consider theories with small Dirac masses. 

One possibility is that the Yukawa coupling $\gamma_D$ in \refl{Higgsdirac} or in a superpotential
term $W_D = \gamma_D L H_u N^c$  is of similar origin but much smaller than  those for quarks and charged leptons. For example,  the lowest allowed $p$ in \refl{Diracmass} could be large, though no concrete models have been 
proposed. Similarly, in a type IIA construction the area of the triangle between the $L, H_u,$ and $N^c$ intersections (see section \ref{suppression})
might be much larger than those for the other couplings due to some anisotropic large internal dimensions. This
is unlikely   for the simplest supersymmetric constructions involving  toroidal orbifolds~\cite{Cvetic:2001nr},
but may be possible for more general manifolds. Similar suppressions involving large anisotropic dimensions
have been found in a heterotic $Z_6$ orbifold~\cite{Ko:2005sh} and in a type I construction~\cite{Ibanez:1998rf,Antusch:2005kf}.

In the following sections, however, we will focus on constructions and models in which the Dirac neutrino  masses are of a different character from those of the other fermions. In many cases, the mechanisms imply
new physics observable at the LHC or relevant to cosmology.

\subsection{Power Law (HDO) or Exponential (D Instanton) Suppressions}\label{diracsuppression}
Barring extreme fine-tuning, a very small Dirac Yukawa coupling suggests that the renormalizable-level perturbative coupling
should vanish. This can be enforced by  a spontaneously broken symmetry of the low energy theory, which might also forbid or suppress Majorana mass terms. Possibilities
include non-anomalous \uprm\ gauge symmetries, global symmetries associated with a family symmetry or anomalous \uprm,
or discrete symmetries such as $Z_n$ (including the possibility of discrete gauge symmetries~\cite{Luhn:2007gq}). 
Tiny Dirac masses may then  be generated by a variety of plausible sub-leading mechanisms.

For example, a low-scale symmetry may forbid a tree-level Dirac Yukawa, but allow a higher-dimensional
operator such as  
\beq W_D= \frac{S L H_u N^c}{\mathcal{M}}, \eeql{wdirac}
where $S$ is a SM singlet, or the analogous Lagrangian term in a non-supersymmetric theory~\cite{Cleaver:1997nj,Langacker:1998ut,Borzumati:2000ya,Murayama:2002je,Chen:2006hn,Demir:2010is}.
This yields a naturally small Dirac mass  $m_D \sim \veva{S} v_u/\mathcal{M}$, which is of the observed magnitude  for $\veva{S}/\mathcal{M}\sim 10^{-12}$.
For $\mathcal{M} =\overline{M}_P$ this implies $\veva{S}\sim 10^3$ TeV, 
as in the \zpr-mediation model~\cite{Langacker:2007ac}.  
Somewhat lower $\mathcal{M}$
would imply  $\veva{S}$ at the TeV scale, suggesting such new physics as a \zpr,
extended Higgs/neutralino sectors, and exotic fermions at the TeV scale~\cite{Barger:2007ay,Langacker:2008yv}. In many cases, the same \veva{S} could lead to an NMSSM-like
dynamical $\mu$ term~\cite{Ellwanger:2009dp}, enhancing the possibility of electroweak
baryogenesis (e.g.,~\cite{Kang:2009rd}). The heavy particles associated
with $\mathcal{M}$ could also lead to Dirac leptogenesis~\cite{Dick:1999je,Murayama:2002je}.

The small Dirac masses may also be associated with supersymmetry breaking. For example, the K\"ahler potential term~\cite{ArkaniHamed:2000bq,Murayama:2004me,Borzumati:2000mc,Borzumati:2000ya,Arnowitt:2003kc,Abel:2004tt}
\beq
K_D=\frac{1}{\mathcal{M}^2} X^\dagger L H_u N^c + h.c.
\eeql{kahlerdirac}
implies $m_D\sim m_{soft} v_u/\mathcal{M}$, which is $\sim 0.1$ eV for $m_{soft} \sim 1$ TeV and $\mathcal{M}\sim 10^{15}$ GeV.
The strong suppressions in non-holomorphic models
are well suited for neutrinos. For example, the K\"ahler term
\beq 
K_{D}\sim \frac{1}{\mathcal{M}^2} X^\dag LN^c\tilde H_d+h.c.
\eeql{del}
leads to a small Dirac masses $m_D\sim m_{soft} v_d/\mathcal{M}$~\cite{Demir:2007dt},
while
\beq 
 K_{D}\sim  \frac{1}{\mathcal{M}} LN^c\tilde H_d+h.c.,
\eeql{bhsv}
found in an F-theory construction~\cite{Bouchard:2009bu}, implies $m_D\sim \mu v_u/\mathcal{M}$.
Other non-holomorphic operators~\cite{Borzumati:2000ya,Demir:2007dt} can lead to trilinear $A\sim m^2_{soft}/\mathcal{M}$ terms for  $\tilde{\nu}_L \tilde{\nu}^c_L H^0_D$, which can lead to a small Dirac mass at loop-level from $\tilde{Z}'$ exchange in a \uprm\ construction.

Small Dirac masses from operators involving quark condensates~\cite{Ibanez:2001nd}, and texture models employing flavor symmetries~\cite{Hagedorn:2005kz}, have also been suggested, as has been the possibility that the $\nu_R$ are actually
composite states from a strongly-coupled sector~\cite{ArkaniHamed:1998pf}.

Perturbative Dirac couplings may be forbidden by anomalous \uprm\  symmetries in intersecting brane constructions,
and then generated by exponentially-suppressed D instantons with  $m_D \sim \exp (-S_{inst})\, v_u$~\cite{Cvetic:2008hi}.
It was shown in a specific construction that the observed mass range can be achieved  naturally, while Majorana masses remain absent.

The generation of small Dirac masses by HDO in the superpotential, in the K\"ahler potential, and by D instantons
is analogous to the solution of the $\mu$ problem~\cite{Kim:1983dt} of the MSSM by NMSSM-like theories~\cite{Ellwanger:2009dp},
by the Giudice-Masiero mechanism~\cite{Giudice:1988yz}, and by D-instantons~\cite{Blumenhagen:2009qh}, respectively.

\subsection{Wave Function Overlaps in Large or Warped Extra Dimensions}\label{large}
There have been many {\em brane-world} models involving large~\cite{ArkaniHamed:1998rs,Dienes:1998vh} or gravitationally warped~\cite{Randall:1999ee} extra dimensions,
differing in the number, sizes, and warping of the extra dimensions, which fields are allowed to propagate in them (the {bulk}) and which are confined to boundaries ({branes}). Such  scenarios can be considered in their own right, or can be motivated by  orbifold GUTs~\cite{Raby:2008gh} involving large internal dimensions, or by the extra dimensions in string theories, especially
those involving a low string scale $M_s$, as in \refl{couplingdefs}. Bulk states would correspond to freely-propagating closed strings, while brane states would correspond to
open strings attached to D branes or to twisted modes stuck at singularities.

Suppose there are $\delta$ compact (flat) dimensions, which we take to have a common length scale $R$ for simplicity,
and internal volume $V_\delta \sim R^\delta$. Comparing the gravitational potential in $3+\delta$ space dimensions for $r\ll R$ with that for $r \gg R$ one finds $\overline{M}^2_P \sim M^{2+\delta}_F V_\delta$,
where $M_F$ is the fundamental gravitational scale in  $4+\delta$ dimensions (which coincides with $M_s$ in \refl{couplingdefs} for $g_s\sim 1$). The Higgs hierarchy problem can be resolved or lessened by choosing
a low $M_F$. For $M_F \sim 10^3$ TeV, for example, one finds $R\sim 10^4,\, 10^{-8},\, 10^{-12}$ cm for
$\delta = 1,2,3$. (Experimental constraints, which clearly exclude $\delta=1$  for that $M_F$,
are reviewed in~\cite{Nakamura:2010zzi}.) The corresponding scales $1/R$ for the Kalzua-Klein
excitations of bulk particles are $\sim 10^{-9},\, 10^3,\,$ and $10^7$ eV.

Small Dirac masses can  emerge naturally if not only gravitons but sterile $\nu_R$ can propagate in the bulk, while the SM particles are confined to the brane~\cite{ArkaniHamed:1998vp,Dienes:1998sb}. 
Expanding the bulk state wave function in Kalzua-Klein modes and evaluating the zeroth mode on the boundary, one
finds a Dirac mass
\beq m_D = \kappa_D\frac{v}{\sqrt{M^{\delta}_F V_\delta}}=  \kappa_D\frac{v M_F}{\overline{M}_P}\sim 0.1  \text{ eV } \frac{ \kappa_D\, M_F}{(10^3\text{ TeV})},
\eeql{leddirac}
independent of $\delta$, where $\kappa_D$ is a dimensionless coefficient.  There are also couplings of $\nu_L$ to the Kaluza-Klein excitations
(larger by \rt), which can also be important in neutrino physics, as well as Dirac masses $j/R$ between (sterile) left-chiral and right-chiral bulk modes. The implications of these small Dirac masses and of the Kaluza-Klein modes, as well as
extensions in which Majorana mass terms are added for the brane and/or bulk states, have been extensively explored.
See, e.g, ~\cite{Antoniadis:2002qm,Davoudiasl:2002fq,Diego:2008zu,Hewett:2004py}, for reviews.

Small Dirac masses can similarly occur in Randall-Sundrum type~\cite{Randall:1999ee}  warped compactifications~\cite{Grossman:1999ra,Gherghetta:2003he} (see~\cite{Chang:2009mv} for a recent discussion). 
These constructions have the interesting feature that they are dual  to theories in which the $\nu_R$ is a composite state in a strongly-coupled sector (cf.~\cite{ArkaniHamed:1998pf})
by the AdS/CFT correspondence~\cite{Maldacena:1997re}.
One can again add small or large Majorana mass terms to these models. Recent discussions include~\cite{Perez:2008ee,Csaki:2008qq,McDonald:2010jm}.

\section{SMALL DIRAC {\em AND} MAJORANA MASSES}\label{smallmixed}

Small Dirac and Majorana masses may both be present. This is suggested by the  LSND/MiniBooNE data
and possibly by some hints in astrophysics/cosmology (see, e.g.,~\cite{Donini:2011jh}).
The LSND/MiniBooNE results could be accounted for if the three light active neutrinos have small mixings of \ord{0.1}\ with one or two light sterile neutrinos, most likely with masses of  \ord{1\text{ eV}}. Neutrino oscillations conserve helicity, so in the context
of  section \ref{basics} the active-sterile oscillations would require that $m_D$ and $m_S$ (and/or $m_T$)
are both small and of roughly the same magnitude. This would preclude the minimal seesaw model\footnote{A hybrid model, involving some light sterile neutrinos and some heavy  ones associated with a seesaw is  of course possible.}
and perhaps suggest that the small masses are associated with the mechanisms described in sections \ref{Majorana} and \ref{Dirac}. Many authors (see, e.g.,~\cite{Donini:2011jh,deGouvea:2006gz}) have suggested the possibility of a {\em mini-seesaw}, e.g., with $m_D\sim 0.2$ eV, $m_S \sim 1$ eV, and $m_T=0$ (or $m_T  \lesssim 0.1$ eV), which leads to the desired order of magnitude of masses and
mixings. 

Such small Dirac and Majorana masses could possibly arise
from unrelated mechanisms. However,  a number of authors have suggested
that the relatively close scales could arise more naturally from higher-dimensional operators, so that, for example,
\beq
m_D \sim \frac{\veva{S}\, v}{\mathcal{M}}, \qquad m_S\sim \frac{\veva{S}^{2}}{\mathcal{M}},
\qquad m_T\sim 0,\eeql{miniscales}
where it is assumed that operators of lower dimension are forbidden by a new symmetry.  
For $\veva{S}\gg v$ one obtains mainly active neutrinos with mass $\sim v^2/\mathcal{M}$,
mainly sterile ones with mass $\sim m_S$, and active-sterile mixings $\sim v/\veva{S}$,
suggesting $\veva{S}=\ord{1 \text{ TeV}}$ and $\veva{S}/\mathcal{M} \sim 10^{-12}$.
As usual, the operators in \refl{opsscales} may have dimensionless coefficients. Furthermore, any multiplicative
symmetry that allows $m_D$ and $m_S$ in \refl{miniscales} would also allow a direct Weinberg operator mass
$m_T\sim v^2/\mathcal{M}$ of the same order of magnitude as the mixing-induced mass,
so the scales motivated by \refl{miniscales} are only approximate. The HDO mini-seesaw has
been motivated by \uprm~\cite{Langacker:1998ut,Chen:2006hn,Sayre:2005yh} or non-abelian symmetries~\cite{Babu:2004mj}, mirror worlds~\cite{Berezhiani:1995yi}, dynamical electroweak symmetry breaking~\cite{Appelquist:2002me}, warped dimensions~\cite{McDonald:2010jm}, and other grounds (e.g.,~\cite{Donini:2011jh}).

Active-sterile mixing requires the simultaneous presence of two distinct types of small mass terms. An alternative to small Dirac and Majorana masses is to introduce Dirac mass terms linking a new left-chiral sterile neutrino with the $\nu_R$, as well as the
conventional active-sterile $m_D$. Something similar occurs in models with large extra dimensions with the  $\nu_R$
and their Kaluza-Klein excitations in the
bulk  (section \ref{large}). These have sterile-sterile  Dirac masses
$j/R$ between left and right chiral Kaluza-Klein modes, which are of suggestive magnitude for $\delta=2$ dimensions 
(e.g., $1/R\sim10$ eV for $M_F\sim 100$ TeV). The mixings are too small in the simplest versions of the model~\cite{Davoudiasl:2002fq}, though they could be enhanced by unequal sizes for the dimensions or by adding additional
mass terms.

Even if eV sterile neutrinos turn out not to be present, there may still be small Majorana masses occurring as
perturbations on dominantly Dirac masses (the pseudo-Dirac limit, $|m_D| \gg |m_{T,S}| \ne 0$).
In particular, Majorana masses are not forbidden by any unbroken gauge symmetry, and exact global symmetries
may be inconsistent with string theory~\cite{Banks:1988yz,Witten:2000dt}. Such perturbations would 
not be important for most purposes. However, even a small splitting between the components of a pseudo-Dirac
neutrino would lead to significant transitions of Solar neutrinos into the sterile states~\cite{deGouvea:2009fp}.
The current observational limits impose the stringent constraint $ |m_{T,S}| \lesssim 10^{-9}$ eV. 
Possible future observations of astrophysical neutrinos from more distant sources would be even more sensitive.

\section{OPERATORS AND SCALES}
\begin{figure}
\centerline{\psfig{figure=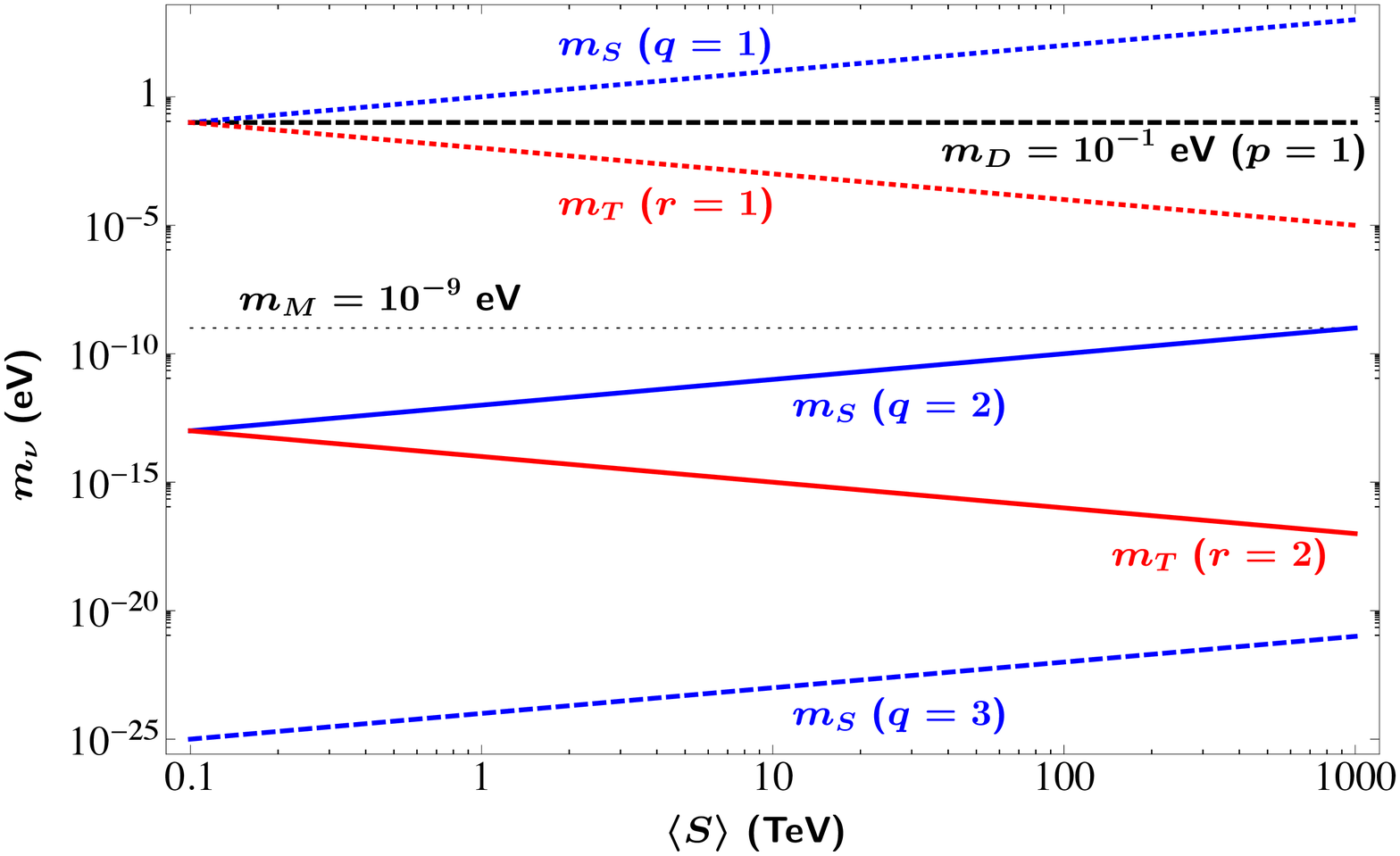,height=20pc}}
\caption{Values of  mass terms for operators of various dimensions as a function of $\veva{S}$,
assuming $\veva{S}/\mathcal{M}\sim 10^{-12}$ (corresponding to $m_D \sim$ 0.1 eV for $p=1$) and $10^{14} \text{ GeV} \le \mathcal{M} \le \overline{M}_P$. The Weinberg operator for $m_T\sim 0.1$ eV corresponds  to $r=1$ and $\mathcal{M}\sim 10^{14}$ GeV. The mini-seesaw corresponds approximately to $p=q=1$, $r>1$, and $\veva{S}\sim $ 1 TeV.  
The pseudo-Dirac limit for $p=1$ has a sufficiently small mass splitting for $m_{T,S} < 10^{-9}$ eV, which holds for all \veva{S}\ for $(q,r) > 1$.
}
\label{operatorscales}
\end{figure}

A number of HDO for $m_D, m_T,$ and $m_S$ have been discussed in the previous sections, involving a variety of
new physics scales $\mathcal{M}$, singlet VEVs \veva{S}, and dimensions. Much of this can be summarized by the generic
HDO-induced masses
\beq
m_D \sim \frac{\veva{S}^p\, v}{\mathcal{M}^p}, \qquad m_S\sim \frac{\veva{S}^{q+1}}{\mathcal{M}^q},
\qquad m_T\sim \frac{\veva{S}^{r-1}v^2}{\mathcal{M}^r},\eeql{opsscales}
where $p,q\ge 0$ and $r \ge 1$ are the lowest dimensions allowed by the symmetries of the theory. 
It should be emphasized that, e.g., $\veva{S}^p$ may be a shorthand for a product of distinct SM singlet VEVs, $\veva{S_1}\cdots \veva{S}_p$, that multiple $\mathcal{M}$'s are possible,  and that dimensionless coefficients have been absorbed into the scales
(e.g., an  $r=1$ seesaw mass  $m_T \sim 0.1$ eV could be induced by a  sterile  $m_S \sim 3$ TeV for a Dirac Yukawa comparable to the $e^-$ Yukawa coupling).
The masses in \refl{opsscales} are shown (for unit coefficients) in Figure \ref{operatorscales} as a function of \veva{S}\
for various $q$ and $r$, for the case $p=1$ with  $\veva{S}/\mathcal{M}\sim 10^{-12}$ (which yields $m_D \sim 0.1$ eV). 


\section{SUMMARY}
A great variety of models of small neutrino mass have been motivated by grand unification or bottom-up
considerations, the most popular being the minimal type I seesaw. String constructions may be very different, however, in part because of possible
additional low energy symmetries and string constraints. Versions of the minimal seesaw
(though typically with non-canonical family structure) are present 
amongst the large landscape of string vacua, though it is not clear how common.
One point of view is to simply focus on the search for such string vacua. However,
another is  to keep an open mind about other possibilities, which may appear
less elegant from the bottom-up point of view but  may occur
frequently in the landscape. This article has surveyed some of the possibilities that have emerged from
concrete string constructions or may be motivated by them. The emphasis was not on the more model-dependent details of family structure, but rather on the classes of mechanisms
(higher-dimensional operators, nonperturbative string instantons, and wave function overlaps in large or warped extra dimensions)
for generating small Majorana and/or Dirac mass scales.

\paragraph{Acknowledgement}I am happy to acknowledge useful discussions and communications
with Mirjam Cveti\v c.
This work was supported by  The Aspen Center for Physics,  NSF grant 1066293, and by the IAS, NSF grant PHYÐ0969448.

\bibliographystyle{arnuke_revised}
\bibliography{pgl_ref}

\end{document}